\documentclass[pdflatex]{article}
\usepackage{waspaa19,amsmath}
\usepackage{graphicx}
\usepackage{bm}
\usepackage{url}
\usepackage{amssymb}
\usepackage{cite}

\newcommand{\argmin}{\mathop{\rm arg~min}\limits}

\title{Batch Uniformization for Minimizing Maximum Anomaly Score of DNN-based Anomaly Detection in Sounds}

\name{Yuma Koizumi${ }^1$, Shoichiro Saito${ }^1$, Masataka Yamaguchi${ }^2$, Shin Murata${ }^1$ and Noboru Harada${ }^1$}
\address{
1: NTT Media Intelligence Laboratories, Tokyo, Japan\\
2: NTT Communication Science Laboratories, Kanagawa, Japan
}

\begin{document}
\ninept
\maketitle

\begin{abstract}
Use of an autoencoder (AE) as a normal model is a state-of-the-art technique for unsupervised-anomaly detection in sounds (ADS). 
The AE is trained to minimize the sample mean of the anomaly score of normal sounds in a mini-batch. 
One problem with this approach is that the anomaly score of rare-normal sounds becomes higher than that of frequent-normal sounds, because the sample mean is strongly affected by frequent- normal samples, resulting in preferentially decreasing the anomaly score of frequent-normal samples. 
To decrease anomaly scores for both frequent- and rare-normal sounds, we propose {\it batch uniformization}, a training method for unsupervised-ADS for minimizing a weighted average of the anomaly score on each sample in a mini-batch. 
We used the reciprocal of the probabilistic density of each sample as the weight, more intuitively, a large weight is given for rare-normal sounds. 
Such a weight works to give a constant anomaly score for both frequent- and rare-normal sounds.
Since the probabilistic density is unknown, we estimate it by using the kernel density estimation on each training mini-batch. 
Verification- and objective-experiments show that the proposed batch uniformization improves the performance of unsupervised-ADS.
\end{abstract}


\vspace{-0pt}


\section{Introduction}
\label{sec:intro}
Since anomalies might indicate mistakes or malicious activities, prompt detection of anomalies may prevent such problems. 
The use of microphones as sensors for anomaly detection, called anomaly detection in sounds (ADS) or acoustic condition monitoring, 
has been adopted in many applications such as audio surveillance \cite{Clavel_2005,Valenzise2007,Ntalampiras_2011,Foggia_2016}, product inspection, and predictive maintenance \cite{Yamashita_2006,Heinicke2015,Koizumi_2017_ADS,sniper}.
In this paper, we specifically consider unsupervised-ADS for product inspection and predictive maintenance. 
Unsupervised-ADS involves detecting ``{\it unknown}'' anomalous sounds by utilizing only given normal sound, 
in contrast to supervised ``Detection and Classification of Acoustic Scenes and Events'' (DCASE) challenge tasks \cite{DCASE2017,DCASE2016_annamaria} for detecting ``{\it defined}'' anomalous sounds such as gunshots \cite{Valenzise2007}.

One of the most common approaches for unsupervised-ADS is outlier-detection \cite{Hodge_2004,Patcha_2007,ASD_survey,DNN_ASD_servey}. 
In outlier-detection, anomalies are detected as patterns in data that do not conform to expected normal behavior. 
The normal behavior is often estimated as a generative model of the observation $\bm{x}$ as $q_{\theta} (\bm{x})$, and the anomaly score of $\bm{x}$ is calculated as its negative-log-likelihood as
\begin{equation}
\mathcal{A}_{\theta}( \bm{x} ) = 
-\ln q_{\theta} (\bm{x}).
\label{eq:anomaly_score}
\end{equation}
Then, $\bm{x}$ is identified as anomalous when $\mathcal{A}_{\theta}( \bm{x} )$ is higher than a pre-defined threshold value $\phi$. 
To train the parameters of the normal model $\theta$ for decreasing $\mathcal{A}_{\theta}( \bm{x} )$ of normal $\bm{x}$, the sample mean of $\mathcal{A}_{\theta}( \bm{x} )$ is minimized:
\begin{align}
\theta
\gets
\argmin_{\theta} \frac{1}{N}\sum_{n=1}^N \mathcal{A}_{\theta}( \bm{x}_n).
\label{eq:ml_train_int}
\end{align}
where $\{ \bm{x}_n \}_{n=1}^N$ denotes a set of samples from the ``true'' probability distribution function (PDF) of normal, $p(\bm{x})$.

One problem in (\ref{eq:ml_train_int}) for unsupervised-ADS is the false-positive (FP) detection of rare-normal sounds. 
Since (\ref{eq:ml_train_int}) is a proximation of the expectation of $\mathcal{A}_{\theta}( \bm{x} )$ on $p(\bm{x})$, $\theta$ is trained so as to preferentially decrease $\mathcal{A}_{\theta}( \bm{x} )$ of frequent-normal sounds. 
Thus, $\mathcal{A}_{\theta}( \bm{x} )$ possibly increases for both unknown anomalous and rare-normal sounds. 
This fact results in frequent FP in our task, because most machine operations consist of several working processes and some processes have extremely shorter operating time than other processes. 
For example, the warm-up time of an engine may be extremely shorter than operating time, thus FPs might be made in every warm-up.

In this paper, we propose {\it batch uniformization}, which is a training method for deep neural network (DNN)-based unsupervised-ADS. 
The key idea of batch uniformization is 
weighting each sample's anomaly score $\mathcal{A}_{\theta}( \bm{x}_n )$ with $w(\bm{x}_n)$ that depends on $p(\bm{x}_n)$, as follows: 
\begin{align}
\theta
\gets
\argmin_{\theta} \frac{1}{N} \sum_{n=1}^N w(\bm{x}_n) \mathcal{A}_{\theta}( \bm{x}_n ).
\label{eq:prop_cost_int}
\end{align}
As a strategy to determine $w(\bm{x}_n)$, we show that the optimal $w(\bm{x}_n)$ is in proportion to the reciprocal of $p(\bm{x}_n)$. 
More intuitively, to obtain small $\mathcal{A}_{\theta}( \bm{x} )$ for both frequent- and rare-normal sounds, a large weight needs to be given for rare-normal sounds. 
By training $q_{\theta} (\bm{x})$ for minimizing such a weighted average, $q_{\theta} (\bm{x})$ is no longer representing $p(\bm{x})$ which is used in traditional outlier-detection;
$q_{\theta} (\bm{x})$ represents the uniform distribution defined on $\{ \bm{x} \mid p(\bm{x}) > 0 \}$, 
thus $q_{\theta} (\bm{x})$ gives a constant $\mathcal{A}_{\theta}( \bm{x} )$ for arbitrary normal $\bm{x}$. 
Since $p( \bm{x} )$ is unknown in practice, 
we estimate it by using the kernel density estimation (KDE) on each training mini-batch.

\section{Conventional method}
\label{sec:conv}

\subsection{Unsupervised anomaly detection in sounds}

ADS is an identification problem of determining whether the state of the target is a normal or an anomaly from the sound emitted from the target. 
Here, we define $\bm{X} = \{ \bm{x}_{t} \in \mathbb{R}^{D} \}_{t=1}^T$ is a time-series of acoustic features extracted from the observed sound. 
Here, $T$ is the number of time-frames corresponding to the operating-time of all operations.

Intuitively, $\mathcal{A}_{\theta}( \bm{X} )$ can be calculated as $-\ln q_{\theta} (\bm{X})$: 
the decision is made by considering all time-frames, {\it i.e.}, batch-type ADS. 
However, it is often unrealistic when $T$ is large for two reasons:
(i) due to the curse of dimensionality, $q_{\theta} (\bm{X})$ becomes harder to estimate, and 
(ii) when an anomalous sound occurs in an early stage of all operations, 
its detection is delayed because the normal/anomaly decision is made at the end of the operation.
Therefore, ``frame-wise ADS'' is often adopted in many studies, that is, the anomaly score is calculated on each frame as 
$\mathcal{A}_{\theta}( \bm{x}_t ) = -\ln q_{\theta} (\bm{x}_t)$, and the normal/anomaly decision is also made for each frame as
\begin{equation}
\mathcal{H}(\bm{x}_t, \phi) =
 \begin{cases}
 0 \mbox{ }(\mbox{Normal}) & \mathcal{A}_{\theta} (\bm{x}_t ) < \phi \\
 1 \mbox{ }(\mbox{Anomaly})& \mathcal{A}_{\theta} (\bm{x}_t ) \ge \phi
 \end{cases},
\label{eq:hard_thres}
\end{equation}
where $\phi$ is a threshold. 
This means that if the anomaly score exceeds $\phi$ even for one frame, $\bm{X}$ is determined to be anomalous.

\subsection{DNN-based unsupervised-ADS}
\label{sec:dnn_conv}

DNNs such as autoencoder (AE) \cite{Tagawa_2015,Marchi_2015,Marchi_2015_IJCNN,Kawaguchi_2017_MLSP}, 
variational AE \cite{VAE_Anomaly,Kawachi_2018,Kawachi_2019}, 
and normalizing flow \cite{adaflow} are used to calculate $\mathcal{A}_{\theta}( \bm{x}_t )$. 
In the case of AE, $\mathcal{A}_{\theta} \left( \bm{x}_t \right)$ is calculated as the reconstruction error
\begin{align}
\mathcal{A}_{\theta} \left( \bm{x}_t \right) = 
\lVert
\bm{x}_t - \mathcal{D}_{\theta_D}( \mathcal{E}_{\theta_E} (\bm{x}_t) )
\rVert _2 ^2,
\label{eq:AE_anomaly_score}
\end{align}
where $\mathcal{E}$ and $\mathcal{D}$ are an encoder and a decoder, respectively, 
$\theta = \{\theta_E , \theta_D \}$ is a set of parameters of an AE, and $\lVert \cdot \rVert_2$ is $L_2$ norm. 
This anomaly score is in proportion to the negative log-likelihood of a Boltzmann distribution as 
\begin{equation}
q_{\theta} (\bm{x}_t) = \mathcal{Z}_{\theta}^{-1} \exp( - \mathcal{A}_{\theta}(\bm{x}_t) ),
\label{eq:AE_pdf}
\end{equation}
where $\mathcal{Z}_{\theta} = \int \exp( - \mathcal{A}_{\theta}(\bm{x}) ) d\bm{x}$ is the normalization constant. 
To minimize $\mathcal{A}_{\theta} \left( \bm{x}_t \right)$ of the normal sounds with respect to $\theta$,
 the sample mean of reconstruction error over samples in a mini-batch $\{ \bm{x}_i^{(u)}\}_{i=1}^{M_u}$ is minimized:
\begin{equation}
\mathcal{J}_{\theta}^{\mbox{\tiny RE}} = 
\frac{1}{M_u} \sum_{i=1}^{M_u} \mathcal{A}_{\theta} \left( \bm{x}_i^{(u)} \right).
\label{eq:ml_train}
\end{equation}

Note that $\mathcal{J}_{\theta}^{\mbox{\tiny RE}}$ does not work so $q_{\theta} (\bm{x}_t)$ becomes close to $p( \bm{x}_t )$, 
because $\mathcal{Z}_{\theta}$ is not incorporated into $\mathcal{J}_{\theta}^{\mbox{\tiny RE}}$ even though $\mathcal{Z}_{\theta}$ is also a function of $\theta$. 
This results in the anomalous sounds also being reconstructed and having small $\mathcal{A}_{\theta} \left( \bm{x}_t \right)$.
To increase $\mathcal{A}_{\theta} \left( \bm{x}_t \right)$ for anomalous sounds, we previously proposed a training method of an AE that works to increase $\mathcal{A}_{\theta} \left( \bm{x}_t \right)$ of simulated anomalous sounds by defining the anomalous sound as ``non-normal'' \cite{Koizumi_2018_IEEE_ADS}. 
As a simplified implementation of our method \cite{Koizumi_2018_IEEE_ADS}, a mini-batch of anomalous sounds $\{ \bm{x}_j^{(a)}\}_{j=1}^{M_a}$ can be generated by adding sounds of other things (hereinafter, something-else sounds) as
\begin{equation}
\bm{x}_n^{(a)} = \bm{x}_n^{(u)} + \alpha \bm{a}_n
\label{eq:anomalous_siml}
\end{equation}
where $n$ is an arbitrary sample index, $\bm{a}_n$ is a something-else sound such as pink-noise, and $\alpha$ is a gain parameter to adjust the anomaly-to-normal-ratio (ANR) \cite{Koizumi_2018_IEEE_ADS}. 
Then, the anomaly score for anomalous sounds can be incorporated into $\mathcal{J}_{\theta}^{\mbox{\tiny RE}}$ as
\begin{align}
\mathcal{J}_{\theta}^{\mbox{\tiny SNP}} &=
\frac{1}{M_u} \sum_{i=1}^{M_u} \mathcal{A}_{\theta} \left( \bm{x}_i^{(u)} \right)
-\mathcal{L}_a,
\label{eq:np_train}\\
\mathcal{L}_a &=
\frac{1}{M_a} \sum_{j=1}^{M_a} \lambda \cdot \tanh \left( 
\lambda^{-1}
\mathcal{A}_{\theta} \left( \bm{x}_j^{(a)} \right) 
\right),
\end{align}
where $\lambda$ is a clipping parameter for avoiding divergence of $\mathcal{A}_{\theta} \left( \bm{x}_t \right)$.

Unfortunately, $\mathcal{J}_{\theta}^{\mbox{\tiny SNP}}$ still has a problem. 
By considering frame-wise ADS from the viewpoint of the batch-type ADS, the definition of the anomaly score is equivalent to
\begin{equation}
\mathcal{A}_{\theta}( \bm{X} ) = 
\max_{ \bm{x}_t } \mathcal{A}_{\theta} \left( \bm{x}_t \right),
\label{eq:def_A_X}
\end{equation}
because if the anomaly score exceeds $\phi$ even for one frame, $\bm{X}$ is determined to be anomalous. 
The maximum value of $\mathcal{A}_{\theta} \left( \bm{x}_t \right)$ tends to depend on $\mathcal{A}_{\theta} \left( \bm{x}_t \right)$ of rare-normal sounds in $\bm{X}$. 
The reason is a part of $\mathcal{J}_{\theta}^{\mbox{\tiny SNP}}$ consists of the average of $\mathcal{A}_{\theta} \left( \bm{x}_t \right)$ on normal samples in a mini-batch, 
and the average is strongly affected by frequent-normal samples.
This results in preferentially decreasing $\mathcal{A}_{\theta} \left( \bm{x}_t \right)$ of frequent-normal samples but does not necessarily decrease that of rare-normal samples.
Therefore, FP detections might be made on all rare-normal sounds such as warm-up sounds.

\section{Proposed method}
\label{sec:prop}

\begin{figure}[ttt] 
\centering 
\includegraphics[width=85mm,clip]
{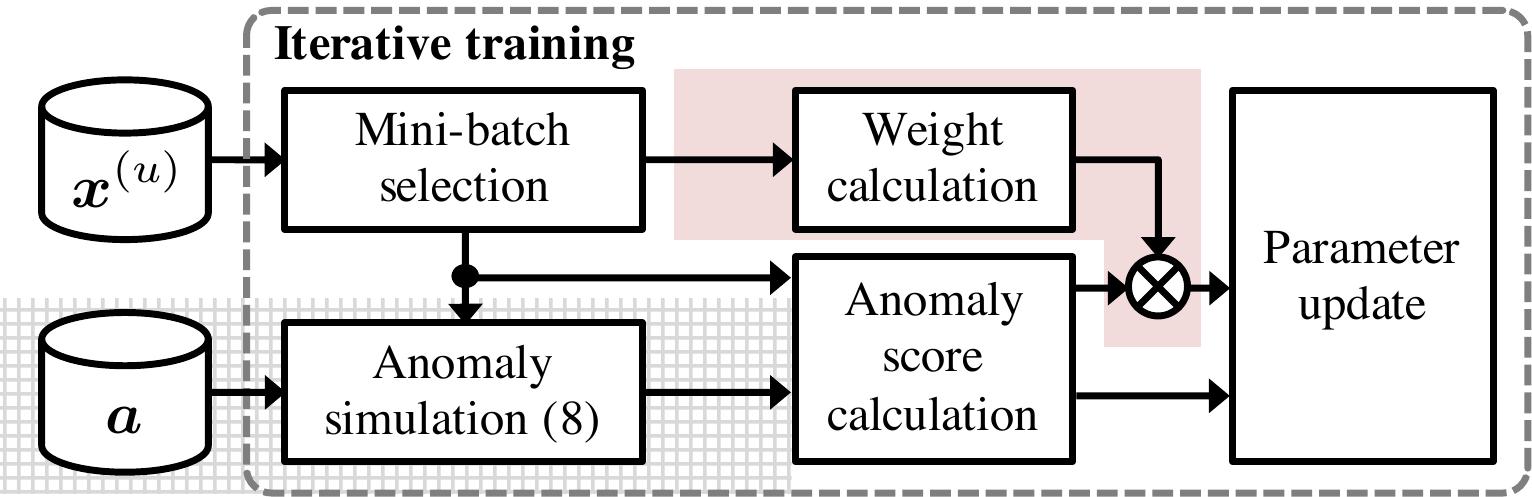} 
\caption{Training procedure of conventional and proposed methods. Gray mesh area is added for $\mathcal{J}_{\theta}^{\mbox{\tiny SNP}}$ and the proposed method $\mathcal{J}_{\theta}^{\mbox{\tiny BU}}$. Pink area is added for the proposed method $\mathcal{J}_{\theta}^{\mbox{\tiny BU}}$. } 
\label{fig:overview}
\end{figure}

Figure \ref{fig:overview} shows an overview of the proposed method, batch uniformization. 
The difference between the conventional and proposed methods is that the proposed method uses the weight calculation. 
In this section, we describe the basic principle and implementation of the weight calculation.

\subsection{Basic principle}

To avoid FP detection, we need to decrease the maximum value of $\mathcal{A}_{\theta} \left( \bm{x}_t \right)$ rather than the average of $\mathcal{A}_{\theta} \left( \bm{x}_t \right)$. 
An intuitive way is directly minimizing the maximum value of $\mathcal{A}_{\theta} \left( \bm{x}_t \right)$ as $\theta \gets \mbox{argmin}_{\theta} \max_{\bm{x}} \mathcal{A}_{\theta}( \bm{x} )$. 
However, this may result in unstable-training because 
the gradient of the cost-function is calculated from only one sample in a mini-batch. 
To stably calculate the gradient, the cost-function should preferably consist of the sample mean.

Here, we consider the normal model whose maximum value of $-\ln q_{\theta}(\bm{x})$ on $\{\bm{x} | p(\bm{x})>0\}$ is the minimum among all PDFs. 
Such a PDF $\mathcal{U}( \bm{x} )$ has a constant probability density on all $\{\bm{x} | p(\bm{x})>0\}$ as
\begin{equation}
\mathcal{U}( \bm{x} ) =
 \begin{cases}
 C 	& p( \bm{x} ) > 0 \\
 0	& p( \bm{x} ) = 0
 \end{cases},
\label{eq:uniform}
\end{equation}
where $C$ is a positive constant that satisfies $\int \mathcal{U}( \bm{x} ) d \bm{x} = 1 $. 
Here we assumed that $p( \bm{x} )$ is a bounded domain distribution.
Therefore, if $q_{\theta}( \bm{x} )$ becomes $\mathcal{U}( \bm{x} )$, $\max_{\bm{x}} \mathcal{A}_{\theta}( \bm{x} )$ is minimized. 
Thus, we consider minimizing the Kullback--Leibler divergence (KLD) between $q_{\theta}( \bm{x} )$ and $\mathcal{U}( \bm{x} )$ rather than actual PDF $p(\bm{x})$ as
\begin{align}
\theta 
\gets \argmin_{\theta} -\int \mathcal{U}( \bm{x} ) \ln q_{\theta} ( \bm{x} ) d\bm{x}.
\label{eq:Uniform_KLD_raw}
\end{align}

\subsection{Batch uniformization using kernel density estimation}

Training with (\ref{eq:Uniform_KLD_raw}) can be realized by roughly two ways:
(i) selecting samples in a mini-batch so that the histogram of the mini-batch becomes uniform, {\it a.k.a.}, mini-batch diversification \cite{bd1,bd2}, or (ii) using weights for each sample that are in proportion to the reciprocal of $p(\bm{x})$ like importance sampling. 
In this study, we adopt the second strategy because mini-batch diversification is difficult to apply to a large-scale dataset. 
By using the second strategy, (\ref{eq:Uniform_KLD_raw}) can be re-written as (\ref{eq:prop_cost_int}) and $w(\bm{x})$ can be calculated as
\begin{equation}
w(\bm{x}) =
 \begin{cases}
 \frac{C}{p( \bm{x} )} 	& p( \bm{x} ) > 0 \\
 0				& p( \bm{x} ) = 0
 \end{cases}.
\label{eq:oracle_weight}
\end{equation}

There are two problems to realize (\ref{eq:prop_cost_int}) with (\ref{eq:oracle_weight}).
The first problem is that the oracle weight (\ref{eq:oracle_weight}) cannot be used because $p(\bm{x})$ is unknown.
In this study, we approximately calculate $p(\bm{x})$ by the kernel density estimation (KDE) as $p ( \bm{x}_i^{(u)} ) \approx \mathcal{K} ( \bm{x}_i^{(u)} )$ from $\{ \bm{x}_i^{(u)}\}_{i=1}^{M_u}$.
For KDE, we use the Gaussian-kernel as
\begin{align}
\mathcal{K} \left( \bm{x}_i^{(u)} \right)
=
\frac{1}{M_u} \sum_{j=1}^{M_u}
\exp
\left\{
-\sigma
\lVert
\bm{x}_{i}^{(u)} -
\bm{x}_{j}^{(u)}
\rVert_2^2
\right\},
\label{eq:KDE}
\end{align}
where $\sigma$ is a band-width parameter. 
The second problem is that $\mathcal{Z}_{\theta}$ is difficult to incorporate into a cost-function when $\mathcal{A}_{\theta} \left( \bm{x}_t \right)$ is calculated as (\ref{eq:AE_anomaly_score}). 
As a tentative solution, we use $\mathcal{L}_a$ in the same manner as $\mathcal{J}_{\theta}^{\mbox{\tiny SNP}} $, and (\ref{eq:prop_cost_int}) can be realized by
\begin{align}
\mathcal{J}_{\theta}^{\mbox{\tiny BU}} =
\frac{1}{\sum_{i=1}^{M_u} w_i } \sum_{i=1}^{M_u} w_i \mathcal{A}_{\theta} \left( \bm{x}_i^{(u)} \right)
-
\mathcal{L}_a,
\label{eq:prop_train}
\end{align}
where $w_i = ( \mathcal{K} ( \bm{x}_i^{(u)} ) + \epsilon)^{-1}$ and $\epsilon$ is a small positive value. 
Since the weights are calculated on each sample in a mini-batch and it may work so that $q_{\theta}( \bm{x} )$ becomes close to $\mathcal{U}( \bm{x} )$, we call the proposed method {\it batch uniformization}.

\section{Experiments}
\label{sec:exp}

We conducted a verification experiment and an objective experiment.
Batch uniformization (BU) was compared with two conventional methods: reconstruction error (RE) and simplified Neyman--Peason cost (SNP) \cite{Koizumi_2018_IEEE_ADS} which are described in Sec. \ref{sec:dnn_conv}. 
Here, BU, RE, and SNP were trained using $\mathcal{J}_{\theta}^{\mbox{\tiny BU}}$, $\mathcal{J}_{\theta}^{\mbox{\tiny RE}}$, and $\mathcal{J}_{\theta}^{\mbox{\tiny SNP}}$, respectively.

\subsection{Verification experiment}
\label{sec:verif}

 \begin{figure}[ttt] 
  \centering 
  \includegraphics[width=87mm,clip]
  {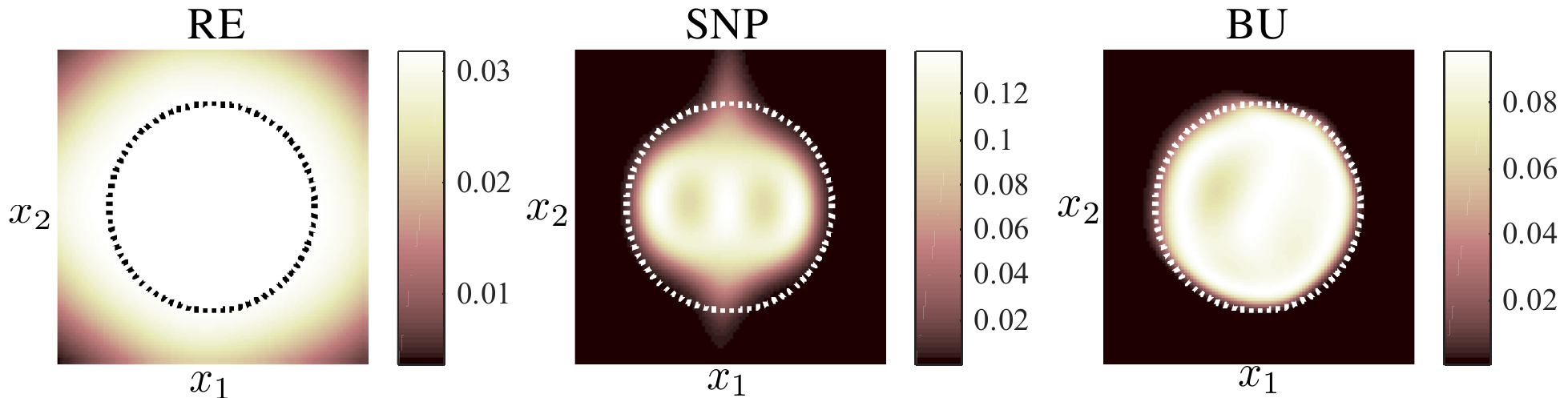} 
  \caption{
PDFs of $\bm{x} = (x_1, x_2)^{\top}$ calculated from AE's anomaly score by (\ref{eq:AE_pdf}) on each grid points in $-3 \leq x_1, x_2, \leq 3$.
The dotted line denotes $r = 2$ (border-line between normal and anomaly).
} 
  \label{fig:estimated_pdf}
 \end{figure}
  \begin{figure}[ttt] 
  \centering 
  \includegraphics[width=90mm,clip]
  {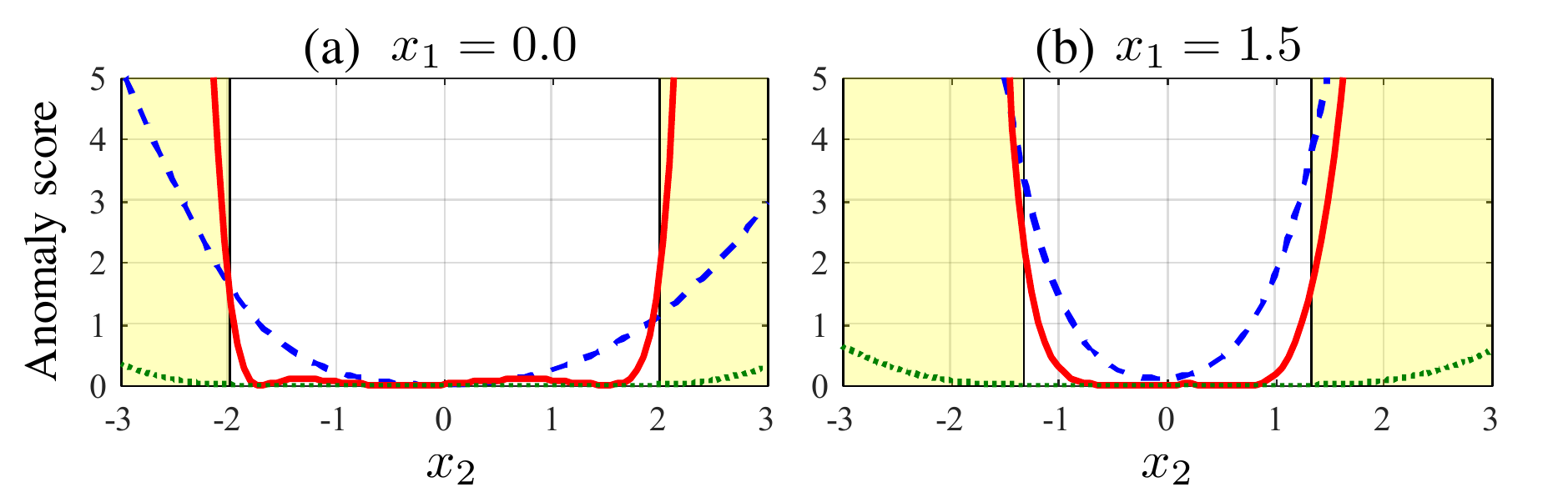} 
  \caption{Anomaly scores of 
  RE (green dotted), 
  SNP (blue dashed), and 
  BU (red solid). 
  Yellow areas denote anomalous areas.
  } 
  \label{fig:anomaly_score}
 \end{figure}
 
 \subsubsection{Dataset and experimental settings}
 \label{sec:setting_velification}

To evaluate whether batch uniformization trains $\theta$ so that $q_{\theta}( \bm{x} )$ becomes close to $\mathcal{U}( \bm{x} )$, we first conducted a verification experiment on an artificial dataset.
The normal and anomalous samples were generated by 
$
\bm{x}_n = r_n \left( \cos(\psi_n), \sin(\psi_n) \right)^{\top},
$
where $r_n$ and $\psi_n$ are the norm and angle parameters, respectively. 
The norms of normal samples were generated as $r_n \sim \mathcal{C}_{[0, 2]}$, 
and that of anomalous samples were generated as $r_n \sim \mathcal{C}_{(2, 3]}$, 
where $\mathcal{C}_{[a, b]}$ is the continuous uniform distribution. 
For both normal and anomalous samples, the angle parameters were generated as $\psi_n \sim \mathcal{C}_{[0, 2\pi)}$. 
Thus, the norm of normal samples was less than or equal to 2, and $p(\bm{x}_n) $ was in proportion to $r_n^{-1}$. 
We generated $N = 10^{4}$ normal and anomalous samples. The minibach size was $M_u = M_a = 500$. 
To evaluate the oracle performance, we used actual anomalous samples instead of simulated anomalous samples by (\ref{eq:anomalous_siml}).

We used an AE consisting of fully-connected-neural-networks (FCN).
The dimensions of each input/hidden/output units were ($2 / 20, 10, 20 / 2$), that is, the AE has 4 FCN layers. 
The activation function of each FCN layer except the output layer was the sigmoid function.
The weight matrices of each layer were initialized by Glorot's method \cite{Glorot_init}, and the AMSgrad \cite{Reddi_2018} with step-size $10^{-3}$ was used as the optimizer. 
The training was stopped after 5,000 updates. The same initial parameters were used for three methods. 
We have decided other parameters as $\sigma = 2D$ so that the Gram-matrix of normal training samples does not become the identity matrix or an ill-condition matrix, and $\lambda = 2D$ based on the tendency of the maximum value of $\mathcal{A}(\bm{x})$ of normal training samples.

\begin{table}[ttt]
  \centering
\caption{KLD between oracle PDFs and estimated PDFs.
$ \mathcal{D} ( p_a(x) \|  p_b (x))$ denotes the KLD between $p_a(x)$ and $p_b(x)$.
}\vspace{2pt}
\small
  \begin{tabular}{c|cc} \hline \hline
Cost-function				& $\mathcal{D} ( p(\bm{x}) \|  q_{\theta} (\bm{x})) $	& $\mathcal{D} ( \mathcal{U}(\bm{x}) \|  q_{\theta} (\bm{x})) $  \\ \hline 
$\mathcal{J}_{\theta}^{\mbox{\tiny RE}}$	& 1.395				& 0.913					 \\ 
$\mathcal{J}_{\theta}^{\mbox{\tiny SNP}}$	& {\bf 0.222}			& 0.333					 \\ 
$\mathcal{J}_{\theta}^{\mbox{\tiny BU}}$	& 0.403				& {\bf 0.083}				\\ \hline \hline
  \end{tabular}\\
  \label{tbl:toy_result}
\end{table}

\subsubsection{Results}

The experimental results are shown in Fig. \ref{fig:estimated_pdf}, Fig. \ref{fig:anomaly_score}, and Table \ref{tbl:toy_result}.
Figures \ref{fig:estimated_pdf} and \ref{fig:anomaly_score} show $q_{\theta} ( \bm{x} )$ and $\mathcal{A}(\bm{x})$, respectively. 
These scores were calculated on each grid point in $-3 \leq x_1, x_2, \leq 3$. Table \ref{tbl:toy_result} shows the KLD from $p(\bm{x})$ and $\mathcal{U}(\bm{x})$. 
The PDF of RE has high probability for both normal and anomaly, thus the anomalous samples were also reconstructed and their anomaly scores became small. 
Meanwhile, in the case of SNP, the probabilities of anomalous samples and KLD between $p(\bm{x})$ and $q_{\theta} (\bm{x})$ were higher and lower than in the case of RE, respectively. 
This means that $\mathcal{L}_a$ successfully works to give high anomaly scores for anomalous data. However, the probability at around $r_n = 2$ was also low, {\it i.e.}, the rare-normal samples. 
That is the problem of minimizing the average anomaly score of normal samples. 
This problem has been solved by the proposed method: as we can see from the PDF of BU, both frequent- and rare-normal samples have high probability. 
The results of the KLD calculation show that the PDF of BU is the closest to the target uniform distribution. 
In addition, Fig. \ref{fig:anomaly_score} shows that the proposed method successfully gave a low anomaly score for normal samples and a high anomaly score for anomalous samples.

\subsection{Objective experiment}
\label{sec:obj_exp}

\subsubsection{Dataset}

To evaluate whether batch uniformization is effective for unsupervised-ADS, we conducted an objective evaluation on a synthetic sound dataset. 
We used a toy-car-running sound dataset in a simulated room of a factory that we used previously \cite{adaflow}: this dataset is freely available on the website\footnote{\scriptsize{\url{https://archive.org/details/toy_car_running_dataset}}}.
This dataset includes four types of car-running sounds and four types of factory noise, and
we generated training/test datasets by mixing
its
at a signal-to-noise ratio (SNR) of 0 dB. 
For the training dataset of something-else sounds $\bm{a}$ used in (\ref{eq:anomalous_siml}), 
we selected wav-files less than 5 sec long from the training dataset of task 2 of DCASE2018 Challenge \cite{DCASE2018task2}. 
All sounds were recorded at a sampling rate of 16 kHz.

The generation process of $\{ \bm{x}_i^{(u)}\}_{i=1}^{M_u}$ and $\{ \bm{x}_j^{(a)}\}_{j=1}^{M_a}$ was as follows. 
Before generating a mini-batch, the whole normal training data was separated so that the length of one wav-file became 3 sec. 
Next, 10 wav-files were randomly selected and concatenated. Then, one something-else wav-file was randomly selected and mixed to the concatenated normal sound at random $-30$ to $10$ dB ANR condition.
Finally, a spectrogram of the mixed sound was calculated, and 
$\bm{x}_t$, which includes a something-else sound, 
were used as $\{ \bm{x}_j^{(a)}\}_{j=1}^{M_a}$, and other $\bm{x}_t$ were used as $\{ \bm{x}_i^{(u)}\}_{i=1}^{M_u}$. 
The definition of the acoustic feature $\bm{x}_t$ is described in the next section. The length of the short-time-Fourier-transform (STFT) and its shift length were 512 and 256 points, respectively.

Since it is difficult to generate various types of anomalous sounds, we created synthetic anomalous sounds in the same manner as in our previous study \cite{Koizumi_2018_IEEE_ADS}. 
A part of the training dataset for task 2 of DCASE2016 Challenge was used as test dataset of anomalous sounds; 140 sounds were selected, including {\it slamming doors }, {\it knocking at doors }, {\it keys put on a table}, {\it keystrokes on a keyboard}, {\it drawers being opened}, {\it pages being turned}, and {\it phones ringing}.
To synthesize the test data, the anomalous sounds were mixed with normal sounds at three types of ANRs: -10, -15, and -20 dB.

\subsubsection{DNN architecture and setup}

We tested four types of AEs: a combination of two types of FCN size and two types of input vector. 
The AEs consists of an encoder and a decoder, and each encoder/decoder has one input FCN layer, $H$ hidden FCN layers, and one output FCN layer. 
Each hidden layer has $U$ hidden units, and the dimension of the encoder output is $Z$. 
The rectified linear unit (ReLU) is used after each FCN layer except the output layer of the decoder.
The parameters of the first size (FCN40/64-large) were $H=4$, $U=512$, and $Z=128$, and those of the second size (FCN40/64-small) were $H=2$, $U=128$, and $Z=40$. 
The input vector $\bm{x}_t$ was defined as
$
\bm{x}_{t} = \left( 
\bm{\psi}_{t, -C}, ...,
\bm{\psi}_{t, C}
\right)^{\top},
$
where $\bm{\psi}_{t, c}= \ln \left[ \mbox{Mel}_M \left[\mbox{Abs} \left[ \left( X_{1,t+c}, ..., X_{\Omega, t+c} \right) \right] \right] \right]$, $X_{\omega, \tau}$ is the STFT spectrum of the observed sound, 
$\omega \in \{ 1,...,\Omega \}$ 
denotes the frequency index, $C$ is the context window size,
and $\mbox{Mel}_M[\cdot]$ and 
$\mbox{Abs}[\cdot]$ denote $M$-dimensional Mel-transform matrix multiplication and the element-wise absolute value. 
Thus, the dimension of $\bm{x}$ was $D = M \times (2C+1) $. The first input vector type (FCN40) used $M=40$ and $C=5$, and the second input vector type (FCN64) used $M=64$ and $C=10$. 
As an implementation for the gradient method, the AMSgrad \cite{Reddi_2018} was used. 
We fix the learning rate for the initial 100 epochs and decrease it linearly between 100--200 epochs down to a factor of 100, where we start with a learning rate of $10^{-4}$. 
We always conclude training after 200 epochs and define an epoch as having 500 mini-batches (about 4 hours). 
The weight matrices of FCNs were initialized by Glorot's method \cite{Glorot_init}. 
We decided other parameters as $\sigma = (2D)^{-1}$ and $\lambda = 2M$ based on the same policy described in Sec.\ref {sec:setting_velification}. 
To implement the KDE, we normalized $\{ \bm{x}_i^{(u)}\}_{i=1}^{M_u}$ before calculating (\ref{eq:KDE}) so that the mean and the variance became 0 and 1, respectively.

\subsubsection{Results}

 \begin{figure}[ttt] 
  \centering 
  \includegraphics[width=86mm,clip]{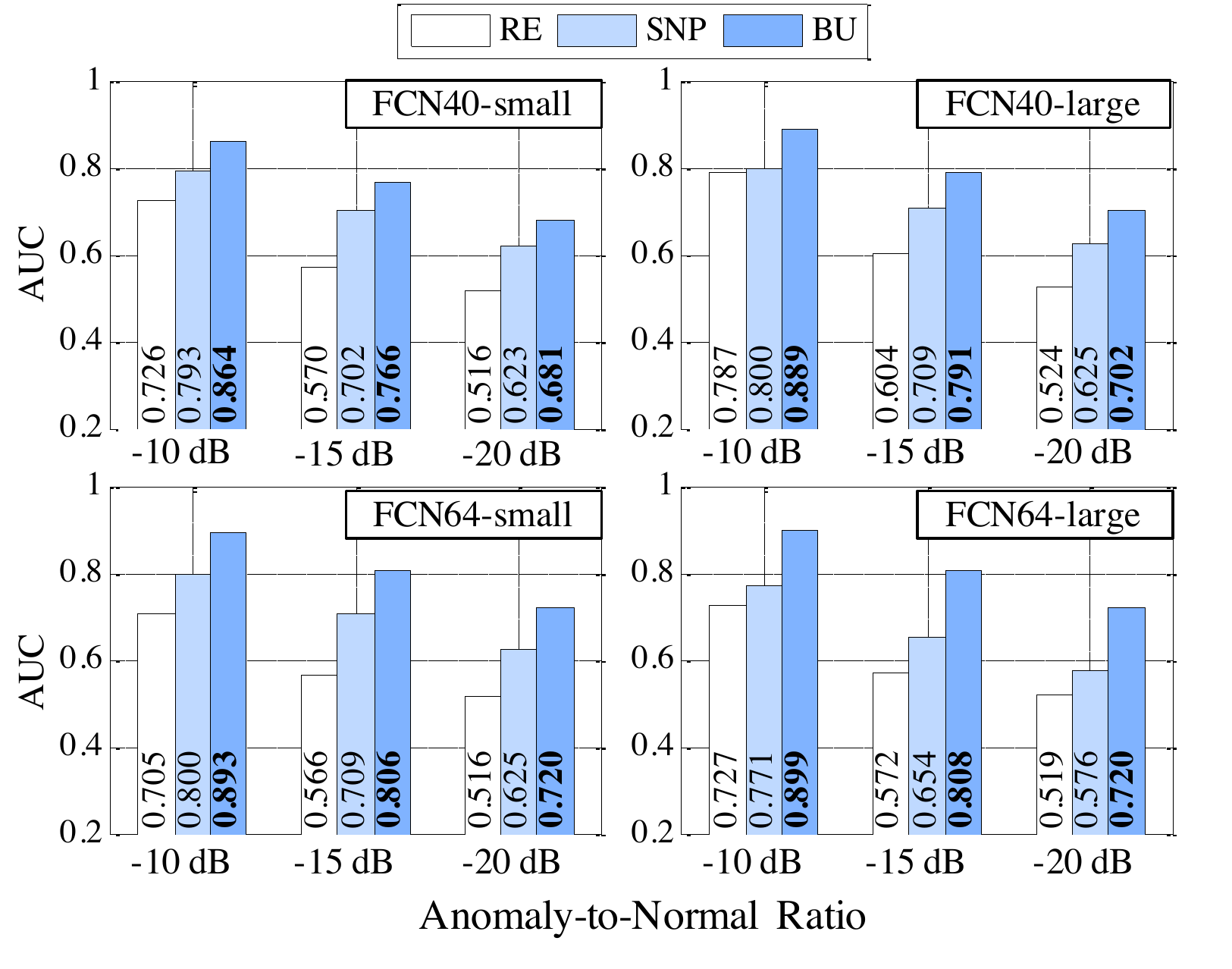} 
  \caption{Evaluation results.} 
  \label{fig:main_result}
 \end{figure}

We used the area under the receiver operating characteristic curve (AUC) as an evaluation metric. 
Figure \ref{fig:main_result} shows the results for AUC on each AEs, ANRs, and cost-functions. 
In all conditions, the proposed method outperformed the conventional methods. 
Since SNP outperformed AE in all conditions, the use of something-else sounds and an additional term to increase the anomaly score of simulated anomalous sounds is effective.
In addition, since the proposed method is an extension of SNP, an AE trained using the proposed method will be effective for unsupervised-ADS.

\section{Conclusions}
\label{sec:cncl}
In this paper, we proposed batch uniformization, a training method for unsupervised- anomaly detection in sounds (ADS). 
The weighted average of the anomaly score was minimized, and the weight was defined as the reciprocal of the probabilistic density of each sample. 
We estimated it by using the kernel density estimation on each mini-batch. 
Verification- and objective-experiments show that the proposed batch uniformization improves the performance of unsupervised-ADS, 
thus, it is effective for deep neural network (DNN)-based unsupervised-ADS.

\clearpage

\end{document}